\documentclass[iop,apj]{emulateapj}
\usepackage{apjfonts}
\usepackage{graphicx}
\usepackage{hyperref}

\newcommand{\kms}{\hbox{$ \, \rm km\, s^{-1}$}} 
\newcommand{\hst}{{\it HST}} 
\newcommand{\hubble}{{\it Hubble Space Telescope}} 
\newcommand{\gaia}{{\it Gaia}} 
\newcommand{\hipparcos}{{\it Hipparcos}}

\begin{document}

\title{A High-Precision Trigonometric Parallax to an Ancient Metal-Poor
  Globular Cluster\altaffilmark{*}}
\shorttitle{Parallax to an Ancient Metal-Poor Cluster}

\author{
T.M. Brown\altaffilmark{1}, 
S. Casertano\altaffilmark{1},
J. Strader\altaffilmark{2}, 
A. Riess\altaffilmark{1,3},\\
D.A. VandenBerg\altaffilmark{4}, 
D.R. Soderblom\altaffilmark{1},
J. Kalirai\altaffilmark{1,3},
and R. Salinas\altaffilmark{5}
}

\altaffiltext{*}{Based on observations made with the NASA/ESA \hubble,
obtained at the Space Telescope Science Institute, which
is operated by the Association of Universities for Research in Astronomy, 
Inc., under NASA contract NAS 5-26555.  These observations are associated
with programs GO-13817, GO-14336, and GO-14773.}

\altaffiltext{1}{
Space Telescope Science Institute, 3700 San Martin Drive,
Baltimore, MD 21218, USA;  tbrown@stsci.edu; stefano@stsci.edu;
drs@stsci.edu; ariess@stsci.edu; jkalirai@stsci.edu}

\altaffiltext{2}{Department of Physics and Astronomy, Michigan State
  University, East Lansing, MI 48824, USA; strader@pa.msu.edu}

\altaffiltext{3}{
  Department of Physics and Astronomy, Johns Hopkins University, Baltimore,
  MD, 21218, USA}

\altaffiltext{4}{Department of Physics and Astronomy, 
University of Victoria, P.O. Box 1700, STN CSC, Victoria, BC, V8W 2Y2, Canada; 
vandenbe@uvic.ca}

\altaffiltext{5}{Gemini Observatory, Casilla 603, La Serena, Chile;
rsalinas@gemini.edu}

\journalinfo{The Astrophysical Journal Letters}
\submitted{Received 2018 February 28; revised 2018 March 7; accepted 2018 March 7; in press}

\begin{abstract}

Using the Wide Field Camera 3 (WFC3) on the \hubble\ (\hst),
we have obtained a direct trigonometric parallax
for the nearest metal-poor globular cluster, NGC~6397.  Although
trigonometric parallaxes have been previously measured for many nearby
open clusters, this is the first parallax for an ancient metal-poor
population -- one that is used as a fundamental template in many
stellar population studies.  This high-precision measurement was
enabled by the \hst/WFC3 spatial-scanning mode, providing
hundreds of astrometric measurements for dozens of stars in the
cluster and also for Galactic field stars along the same sightline.
We find a parallax of 0.418$\pm0.013\pm0.018$~mas (statistical,
systematic), corresponding to a true distance modulus
of $11.89 \pm 0.07 \pm 0.09$~mag (2.39$\pm0.07\pm0.10$~kpc).
The $V$ luminosity at the stellar main sequence turnoff
implies an absolute cluster age of 13.4$\pm 0.7 \pm 1.2$~Gyr.

\end{abstract}


\section{Introduction}

Stellar populations at all redshifts are interpreted within the
framework of stellar evolution models, and
the observational foundation for such models are
Galactic star clusters, because they provide
samples at nearly fixed distance, age, and chemical
composition.  When employing isochrones to interpret 
a stellar population, it is common to cite 
cluster reference points, quoting the ages and metallicities for
which the isochrone library best matches observed color-magnitude
diagrams (CMDs).  For example, the ancient metal-poor anchor in the 
Bruzual \& Charlot (2003) stellar population models is NGC~6397.

There are dozens of open clusters within 1~kpc (Dias et al.\ 2002),
and many have direct parallaxes (e.g., Soderblom et al.\ 2005;
van Leeuwen 2009; van Leeuwen et al.\ 2017), but the nearest globular
clusters are at larger distances that put them beyond the reach
of \hipparcos\ or the Fine Guidance Sensor on the \hubble\
(\hst).  This problem has weakened the observational
foundation for much of astronomy, because the distances to all
metal-poor ([Fe/H]~$< -1$) and ancient (age $>$ 10~Gyr) star clusters
have until now been based upon indirect methods, such as main-sequence subdwarf
fitting (e.g., Gratton et al.\ 2003), RR~Lyrae (e.g., Cacciari \&
Clementini 2003), dynamical modeling (e.g., van der Marel \& Anderson
2010), and white-dwarf (WD) fitting (e.g., Hansen et al.\ 2007).
Distances to fiducial globular clusters are the largest uncertainty
when using them to anchor stellar models.

That situation is now on the verge of dramatic
improvement. With the advent of spatial scanning (MacKenty 2012; Riess
et al.\ 2014; Casertano et al.\ 2016), the nearest globular
clusters are within reach of \hst, and many more are within reach
of \gaia\ (Pancino et al.\ 2017).  The closest is M4 -- an
intermediate-metallicity ([Fe/H]~=~$-1.15$; Kraft \& Ivans
2003) cluster with an age of 11.5~Gyr (VandenBerg et
al.\ 2013) and highly uncertain distance (1.7--2.2~kpc; e.g., Harris 1996;
Hansen et al.\ 2004; Bedin et al.\ 2009),
due to its unusual foreground reddening.  
NGC~6397 is the next closest, and a much better template for ancient
metal-poor populations.  Its distance modulus has been determined
by both main-sequence fitting (12.13~mag, Reid \& Gizis 1998;
12.01~mag, Gratton et al.\ 2003) and WD fitting
(12.03~mag, Hansen et al.\ 2007), implying a distance of
2.6~kpc.  Independent spectroscopic metallicity measurements give
[Fe/H]~=~$-2.03$ (Gratton et al.\ 2003) and $-2.02$
(Kraft \& Ivans 2003), and fitting of the main-sequence turnoff (MSTO)
implies an age of 13.0~Gyr (VandenBerg et al.\ 2013).  Its
Galactic latitude ($l$~=~338.17, $b$~=~$-11.96$) facilitates spatial
scanning of both cluster and reference field
stars of suitable brightness and distance.  It is moderately reddened;
we assume $E(B-V)$~=~0.185~mag, based upon measurements of
0.183~mag (Gratton et al.\ 2003), 0.186~mag (Schlegel et al.\ 1998),
and 0.187~mag (Anthony-Twarog et al.\ 1992).  NGC~6397 is of moderate
luminosity ($M_{Vt} = -6.64$~mag; Harris 1996), and thus much
less massive than clusters with complex populations such as NGC~2808 and
$\omega$~Cen (e.g., Piotto et al.\ 2015).

\begin{figure*}[t]
\begin{center}
\includegraphics[width=6.0in]{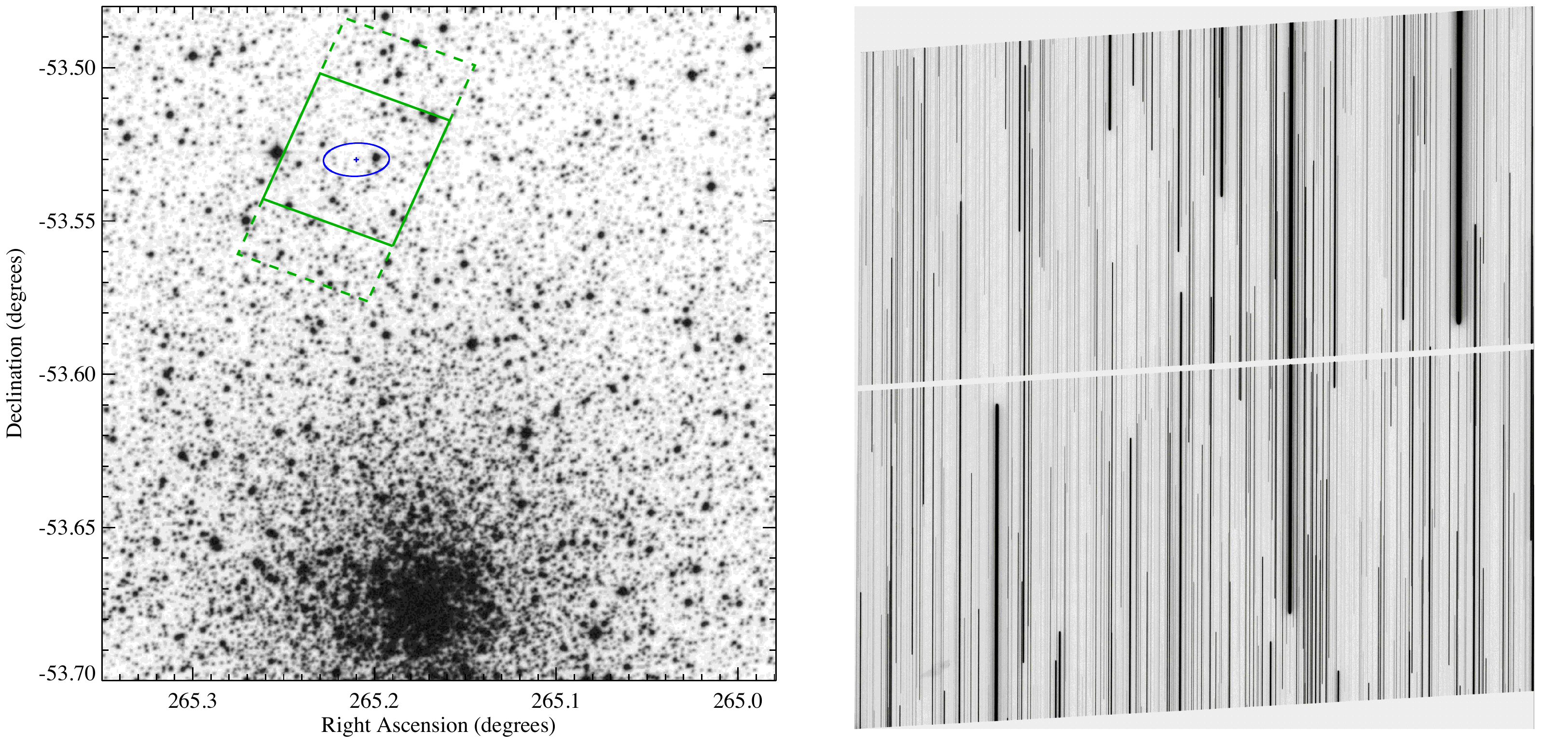}
\end{center}
\caption{Left panel: a Digital Sky Survey image of NGC~6397
  showing a simplified WFC3 footprint (solid green) at the
  center of the area scanned (dashed green). The area and
  position angle were chosen to maximize the number of bright ($V <
  18$~mag) field and cluster stars that could be scanned without
  neighbor contamination, while maximizing the parallax motion that can be
  measured within the observing constraints.
  The parallax ellipse for stars in NGC~6397 (blue)
  is shown magnified by 10$^5$.
  Our astrometric information is in a single axis,
  perpendicular to the scan.  Right panel: a scan of
  the NGC~6397 field, shown at a clipped linear stretch.}
\end{figure*}

Because of its fundamental importance as a population template, we
have obtained \hst\ spatial scans of NGC~6397 with
the goal of obtaining a high-precision measurement of its
trigonometric parallax.  We are currently achieving a parallax
precision $\sim$20--100~$\mu$as per star, which is competitive with
the measurements that will be obtained by \gaia.
In \gaia\ Data Release 1, parallaxes for nearby globular clusters
are barely detectable (Watkins \& van der Marel 2017).
Single-star
precisions in the upcoming \gaia\ data release are likely to be at
the level of 100~$\mu$as (Prusti 2012), and possibly somewhat better.
Compared to our measurements, the final \gaia\ parallaxes will
have smaller uncertainties for the brightest M giants, and
larger uncertainties at $V\gtrsim$~14~mag.  \gaia\ results rely
upon exquisite orientation knowledge for widely separated fields, while
\hst\ results are based upon precise relative measurements
anchored to field reference stars.  Given the distinct systematics,
our measurement provides an independent check of 
\gaia\ measurements for this critical cluster.  Presently,
our uncertainties are dominated by systematic errors,
but we continue to improve our analysis, and plan to address
these in a future paper.  Given our current precision,
using a method completely independent from previous distance estimates
and upcoming \gaia\ results, it is appropriate to provide a preliminary
result at this time.

\section{Data}

\subsection{\hst\ Observations}

We obtained five orbits of WFC3 imaging over the course of 2~years,
with one orbit every six months, beginning in 2014 September, 
timed to occur near the maximum
parallactic motion of NGC~6397.
A link to the \hst\ data is provided here:
  \href{http://dx.doi.org/10.17909/T9SX1F}{10.17909/T9SX1F}.
Each epoch included four spatial scans in the F606W
filter and eight direct images in the F336W, F467M, F547M, and F850LP
filters.  The spatial scans did not hold a fixed position, but instead
trailed the field at
0\farcs41~s$^{-1}$ across the detector for 3600~pix along a path that
produced trails approximately aligned with the y-axis (Figure~1), deviating
by $\sim$0\fdg05 to
provide sub-sampling of the point spread function (PSF) perpendicular
to the scan.  By trailing the image, we obtained hundreds of
astrometric measurements for each star, while drastically reducing
systematics from geometric distortion, jitter, and PSF sampling,
but the astrometric information is only available in a direction
perpendicular to the trails.  Ideally, we would have scanned the
field in a direction that was perpendicular to the long axis
of the parallax ellipse at the time of greatest parallactic offset,
but a tradeoff must be made between available telescope roll, parallax
motion, and overlapping trails between stars.  The
optimized solution gave observations at the desired date but rotated
by 27\fdg6 from the ideal position angle, reducing the measurement
sensitivity by a factor of 0.89.  At
this orientation, our measurements of the motions for cluster stars
are more tangential than radial (57\fdg4 from radial).

The spatial scans produced useful trails for stars
brighter than $V\lesssim$~18.5~mag.  To characterize the stars
associated with each trail requires a photometric catalog
reaching at least this depth, and the \hst\ direct images are far
deeper.  Given the sparse field, PSF-fitting photometry is
unnecessary, and we produced catalogs from the direct images using
aperture photometry (0\farcs2376 radius) derived with the {\sc aper}
routine (Landsman 1993).  The \hst\ catalogs were then merged with
ground-based Str\"{o}mgren photometry of the same field
(Anthony-Twarog \& Twarog 2000), which also reaches well below $V =
18.5$~mag.  Using the same criteria as Anthony-Twarog \& Twarog
(2000), based upon the full Str\"{o}mgren photometry, we characterized
stars as cluster members or field stars (see Figure~2).

\subsection{Southern Astrophysical Spectrograph (SOAR) Observations}

Our \hst\ spatial scans yield relative parallaxes,
and these must be put in an absolute frame using distance
estimates for the field stars along the sightline.  The
distance estimates are primarily based upon multi-band photometry
of our field, but to supplement this information for a subset of
stars, we obtained multi-object spectroscopy with the Goodman
Spectrograph (Clemens et al.\ 2004) on the 
SOAR telescope, with a wavelength coverage of
$\sim$350--580~nm.  In practice, this yielded spectral classifications
for only 14 of the 89 field stars used to determine the
parallax absolute frame.  For most of these 14 stars, the spectra
reaffirmed the photometric characterization, but for a
few, the spectroscopy allowed us to distinguish between multiple
possibilities (i.e., dwarfs vs.\ giants).

\begin{figure}[t]
\begin{center}
\includegraphics[width=3.25in]{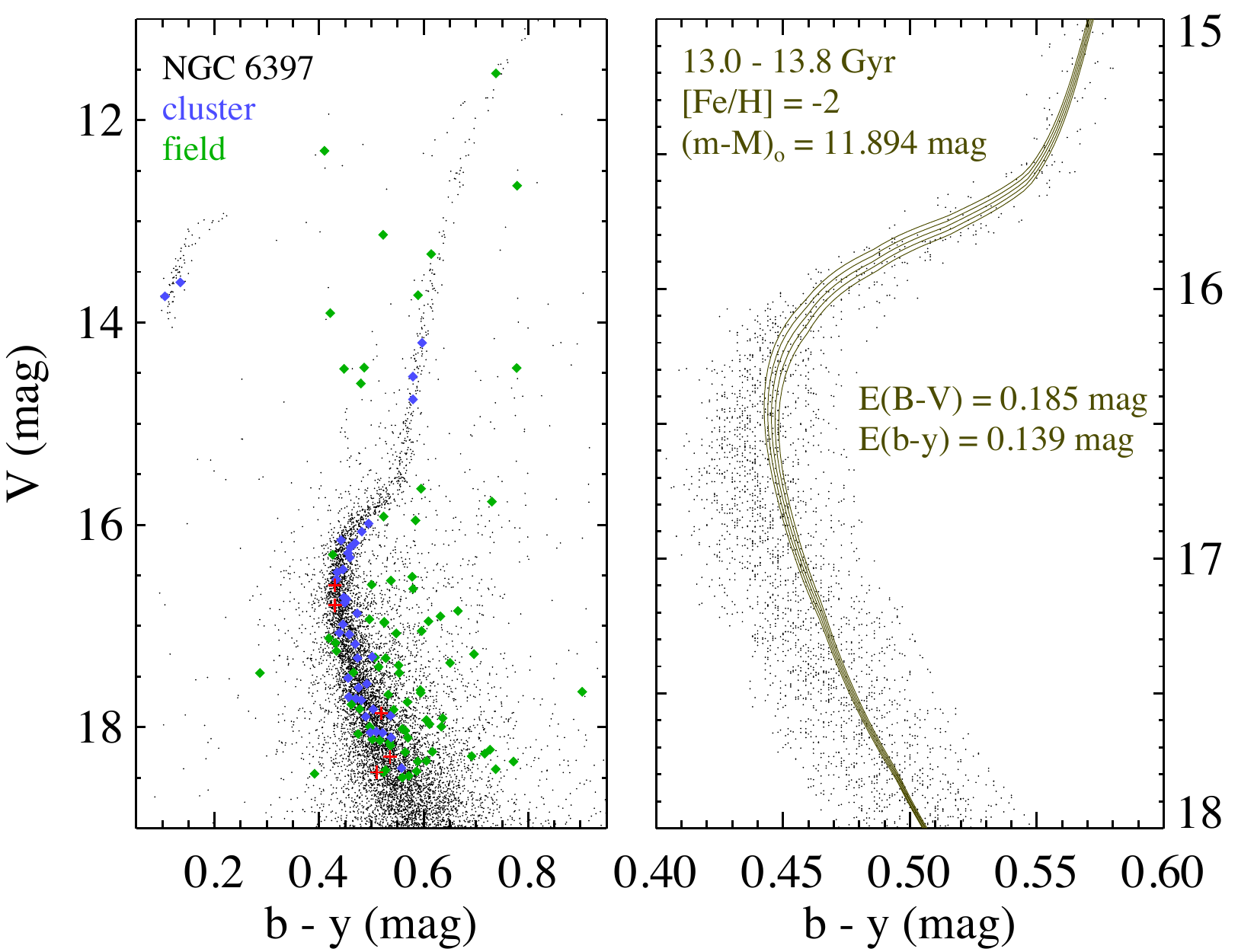}
\end{center}
\caption{Left panel: the CMD for NGC~6397 (Anthony-Twarog \&
  Twarog 2000).  Stars with clean trails in the \hst\ spatial
  scans are indicated in blue (cluster) and green (field), using the
  same membership criteria employed by Anthony-Twarog \& Twarog
  (2000), based upon the full set of Str\"{o}mgren photometry and
  indices.  Red crosses indicate stars included in our cluster sample
  that appear to be outliers (see the text and Figure~4).  Right
  panel: an expansion of the CMD at the MSTO, which is sensitive to
  age, showing only cluster members as designated in Anthony-Twarog \&
  Twarog (2000).  Isochrones (VandenBerg et al.\ 2014) are shown for a
  range of ages at the cluster metallicity ([Fe/H]~=~$-2.0$,
  [$\alpha$/Fe]~=~$+0.4$), the assumed extinction, and our derived
  distance modulus (labeled), with the $b-y$ color calibrated to match
  the base of the red giant branch. The locus is best matched by the
  13.4~Gyr isochrone.}
\end{figure} 

\section{Analysis}

The use of \hst\ spatial scans to measure parallaxes
was pioneered in Cepheid observations (e.g., Riess et al.\ 2014;
Casertano et al.\ 2016).  The technique is summarized in Riess et
al.\ (2018 and references therein).
Here, we briefly summarize the technique,
along with the distinctions between our program and
the Cepheid programs.

\begin{figure*}[t]
\begin{center}
\includegraphics[width=6.5in]{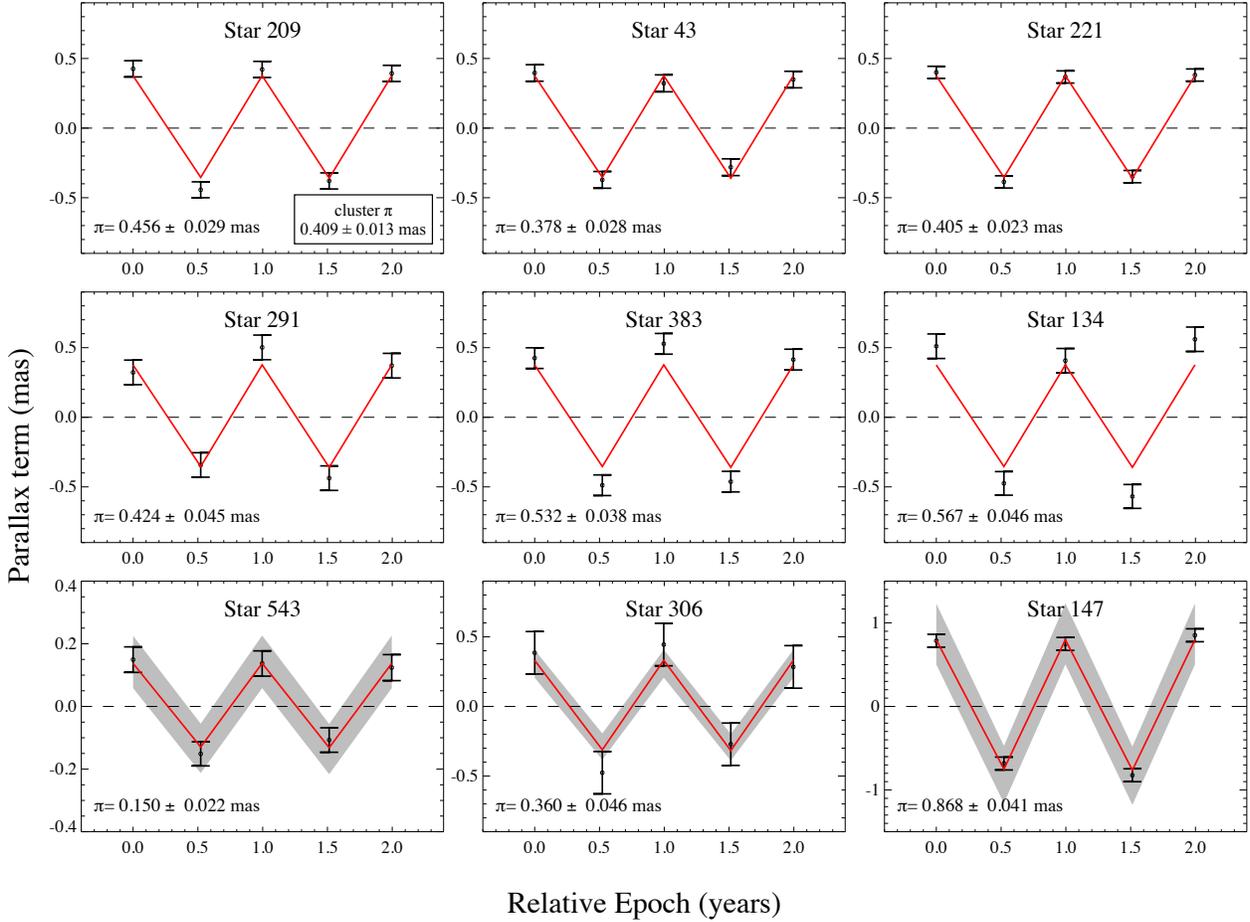}
\end{center}
\caption{Top six panels: the proper motion subtracted 1D motions
  for six representative cluster stars, as observed
  over five epochs (2~years).  These motions come from our 
  analysis of the full cluster sample of 44 stars
  (left panel in Figure 4).  The red line indicates the
  measured cluster parallax from the full solution that assumes all
  of the cluster stars are at the same distance, with all errors
  propagated.  The individual parallaxes
  represent the pure astrometric measurement for each star in isolation.  
  Bottom three panels: the same, but for three
  representative field stars from our sample of 89.
  The red line indicates
  the measured parallax for each field star,
  while the gray band indicates the photometric
  parallax with 2$\sigma$ width.}
\end{figure*}

Each spatial scan produces trails that
are nearly aligned with detector $Y$.  Astrometric
measurements are made along detector $X$ by fitting a
position-dependent line-spread function to each 15 pixel minirow
along a scan, excluding cosmic rays and detector artifacts, giving
many measurements of $X$ as a function of $Y$ for each
star.  These relative detector coordinates are transformed to relative
sky coordinates using a geometric distortion solution determined as in
Bellini et al.\ (2011), including an empirically derived
delta geometric distortion map (Casertano et al.\ 2016),
with corrections for time-dependent
plate-scale variations, frame-to-frame rotation, and rotation during
an individual scan.  The relative astrometry is registered using
a 2D second-order polynomial that accounts for time-dependent distortions
along the measurement direction, and
measured relative to a reference line that contains the jitter
history, length, and slope of the scan, constructed from the
superposition of all time-aligned scan lines.  A model is then fit
simultaneously to the cluster and reference field stars,
using for the latter photometric distance constraints derived by
comparing our multi-band photometry to the stellar population along
this sightline in the Besan\c{c}on Model of the Galaxy (Robin et
al.\ 2003).  Synthetic Str\"{o}mgren and \hst\ photometry for the stars
in the Besan\c{c}on model was calculated
using MARCS spectra (Gustafsson et
al.\ 2008), supplemented by Castelli \& Kurucz (2003)
spectra at high temperatures.  The stellar motions
are modeled as the superposition of a relative proper motion
and parallax over five epochs (2~years), accounting
for the projection of the parallax ellipse on detector $X$.
In Figure~3, we show the parallax motions
for representative cluster and field
stars, as derived in one of the two bounding solutions discussed below.

\begin{figure*}[t]
\begin{center}
\includegraphics[width=5.0in]{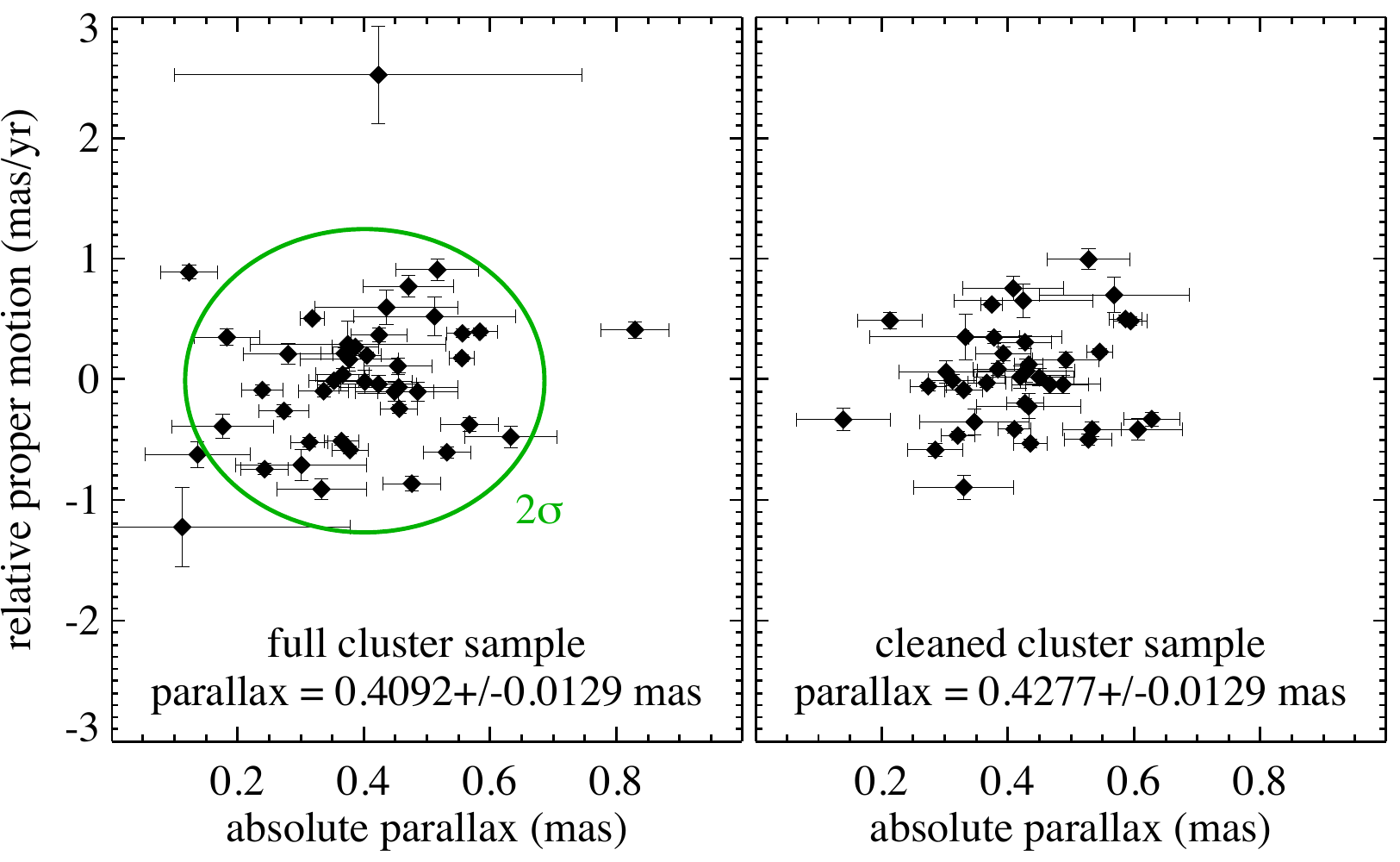}
\end{center}
\caption{Left panel:
  individual measurements of proper motion and
  parallax for our full sample of 44 cluster stars (see Figure~2).
  For comparison, the ellipse axes give the 2$\sigma$
  variation in the measurements of proper motion and parallax, centered
  on the median value of each.  Five stars fall outside of this ellipse.
  The parallax from this full sample is 0.4092~mas.  The weighted
  average of the measurements here is not significantly changed if
  the outliers are removed.  Note that the cluster
  parallax in the full solution propagates all errors, such that the
  cluster parallax uncertainty is larger than that one would expect
  from a weighted average of the individual measurements.
  Right panel: the full solution derived from 39 stars after removal
  of the five outliers.  The result is an increased
  cluster parallax of 0.4277~mas, and the individual measurements have
  appreciably shifted with respect to the previous solution.}
\end{figure*}

Our program has several distinctions from the Cepheid programs.
First, in the Cepheid programs, each image targets a single
bright Cepheid and dozens of relatively faint field stars,
necessitating the use of multiple filters to obtain the appropriate signal;
the cross-registration of these distinct filters incurs a
cost in the error budget.  Here, each image targets multiple cluster
stars of similar brightness to the field reference stars (see
Figure~2), conveying the advantage that all scans are in a
single filter (F606W).  Second, in the Cepheid analysis, the reference
frame is derived by comparing the field stars to a set of isochrones
anchored to the Besan\c{c}on model, 
providing a finer grid of stellar parameters than the model itself.
Here, we have translated the Besan\c{c}on model directly to synthetic
\hst\ and Str\"{o}mgren photometry for comparison to
the observations. The comparison to the Besan\c{c}on model
produces a probability distribution function (PDF) for the estimated
distance to each field star.  Where the PDF has multiple
peaks, the first iteration of the analysis uses a prior spanning these
peaks, and then subsequent iterations down-select to a single peak
consistent with the previous solution.  Third, the
Cepheid analysis iteratively solves for the extinction along the
sightline, using the Schlafly \& Finkbeiner (2011)
dust extinction map as a prior.  Here, we rescale the
extinctions in the Besan\c{c}on model such that $E(B-V)$~=~0.185~mag
at the distance of NGC~6397, given the foundation of
literature associated with the cluster; otherwise, the
Besan\c{c}on model would give a much lower value (0.141~mag) for this
sightline at the cluster distance.
In comparison to the Cepheids, the higher Galactic latitude of NGC~6397
puts nearly all of the stars in our sample outside of the thin disk,
imposing essentially constant extinction.
Fourth, the Cepheid programs have
sampled 4--9 epochs per star, while here we have sampled 5 epochs.
Fifth, we adopt a weak proper motion prior for the cluster stars, with
a width of 15~\kms\ in the resolution direction, and a mean proper motion
of zero (we have no absolute proper motion reference).  This prior
constrains the inter-epoch alignment of the overall solution, but
our derived proper motions and parallaxes come solely from our astrometric
measurements.  No such prior exists in the Cepheid programs.
Finally, the Cepheid program images each have a single Cepheid,
while our images have multiple cluster stars
assumed to be at the same distance; we thus apply
weights to account for fit quality when
combining the individual cluster
measurements into a single cluster distance.  Specifically, we
increase the errors for each cluster star such that $\chi^2 < 2$, and
then weight each star in inverse proportion to its errors squared.

The level of precision we are seeking in our program is in a different
regime than that in the Cepheid programs, making it sensitive to
effects at the $\sim$2$\times$10$^{-4}$~pix level, compared to
$\sim$10$^{-3}$~pix in the Cepheid programs.  At this precision, our
program has revealed a sensitivity to systematic uncertainties that
currently dominate our analysis.  We defer a more complete discussion
of these systematics to a full-length paper (S.\ Casertano et al.\ 2018,
in preparation).  Here we note that the solution changes significantly
depending upon the samples employed in the analysis.  Two solutions
that bound the results are shown in Figure~4.  In the left panel, the
relative proper motion and astrometric parallax are shown for each of
the 44 stars in our cluster sample.  Using the full sample, the
cluster parallax is 0.409~mas, and the one-dimensional velocity
dispersion is 6.2~\kms, although we note that the direction of our
measurement is largely tangential (see Figure~1).  There are several
apparent outliers, despite the fact that they appear to be members in
the Str\"{o}mgren and \hst\ photometry.  If we median clip the sample
at 2$\sigma$ (green circle), this removes five outliers.  The weighted
average of the remaining 39 measurements, without re-running the full
solution, gives a parallax of 0.411~mas, which is not significantly
distinct from that in the full sample, although the velocity
dispersion drops to 5.0~\kms.  For comparison, the central velocity
dispersion of NGC~6397 has been measured at 4.5~\kms, dropping to
2.2~\kms\ in the outskirts (Meylan \& Mayor 1991; see also Kamann et
al.\ 2016).  However, if we then reprocess the full solution with that
reduced sample of 39 stars (right panel), the parallax increases to
0.428~mas -- a change larger than that expected from the statistical
errors or from the actual values of the individual parallaxes
themselves.  The velocity dispersion in this new solution is 4.5~\kms.
These five outliers may be stars with undetected problems (e.g.,
contamination by a fainter star, unresolved binary), or they may not
be true cluster members.  A subset might be legitimate measurements of
cluster stars, because it is not unreasonable for one or two stars to
scatter beyond 2$\sigma$. Our preliminary result takes the midpoint of
these two bounding cases, giving a cluster parallax of 0.418~mas, with a
systematic uncertainty of 0.010~mas due to the dependency on the
sample, and a statistical uncertainty of 0.013~mas, dominated by the
correction to absolute parallax.

There is good agreement between our photometric priors and astrometric
results for the field population, with half the sample agreeing at the
1$\sigma$ level, and 70\% of the sample agreeing at the 2$\sigma$
level.  However, there is an additional systematic uncertainty
associated with the photometric parallax priors of the field reference
stars, due to the extinction uncertainty along this sightline and the
uncertainties in the absolute luminosities for the model stellar
population.  We can estimate this systematic error by shifting the
distance moduli of the reference stars by $+$0.1~mag, which will have a
larger impact on the parallax priors for the relatively nearby stars.
For such a shift,
the cluster parallax solution changes by $-$0.015~mas.  Adding
these systematics in quadrature, our systematic
uncertainty for the cluster parallax increases to 0.018~mas.  The true
distance modulus is $11.89 \pm 0.07 \pm 0.09$~mag ($2.39 \pm 0.07 \pm
0.10$~kpc)

\section{Discussion}

We have measured a trigonometric parallax of $0.418 \pm 0.013 \pm
0.018$~mas for the nearest metal-poor globular cluster, NGC~6397.  The
distance modulus, $11.89 \pm 0.07 \pm 0.09$~mag, is shorter
than most measured previously, at the level of 1--2$\sigma$ (cf.\ Reid
\& Gizis 1998; Gratton et al.\ 2003; Hansen et al.\ 2007), but
consistent with a recent dynamical distance (Watkins et al.\ 2015).  We
note that new subdwarf fits to NGC~6397 also yield a relatively short distance
(D.\ VandenBerg et al.\ 2018, in preparation) that is consistent with our
results.

With the upcoming catalog from \gaia\ Data Release 2, we will have an
independent measurement of the NGC~6397 distance for comparison, to
the extent that the \gaia\ results for individual cluster
stars can be combined, accounting for the correlated errors on the
scale of the cluster.  Furthermore, the \gaia\ results can be used
to augment the \hst\ analysis by redefining the distance priors
for the reference field stars.  The combined measurement
should yield a parallax uncertainty at the level of $\sim$1\%.

Relative cluster ages are generally measured in two ways: the
luminosity difference between the horizontal branch and the MSTO
increases at older ages, and the color difference between the MSTO and
the base of the red giant branch decreases at older ages.  Combining
these methods, VandenBerg et al.\ (2013) found an age of 13.0~Gyr for
NGC~6397.  The MSTO absolute luminosity is also an
absolute age indicator, when compared to isochrones at fixed
distance.  In Figure~2, we show the VandenBerg et al.\ (2014)
isochrones translated into the frame of the Anthony-Twarog \& Twarog
(2000) photometry, using our assumed extinction and derived
distance.  The $V$ luminosities are those in the VandenBerg et
al.\ (2014) library; the $b-y$ colors come from synthetic
photometry, using MARCS spectra (Gustafsson et al.\ 2008) and the
temperatures and gravities along the isochrones.  The $b-y$
transformation has been calibrated to retain the MSTO
color, but match the observed color difference between the MSTO
and the base of the red giant branch.  The comparison to the
isochrones implies an absolute cluster age of 13.4~Gyr.

There are significant statistical and systematic
uncertainties when deriving an absolute age.  The MSTO absolute
luminosity changes by $\sim$0.1~mag per Gyr at old ages, such that the
statistical (0.07~mag) and systematic (0.09~mag) uncertainties in
distance modulus correspond to 0.7~Gyr and 0.9~Gyr in age.  Modern
isochrone libraries (e.g., VandenBerg et al.\ 2014; Choi et al.\ 2016;
Marigo et al.\ 2017) agree at the level of $\sim$0.06~mag for the
absolute MSTO luminosity at a particular chemical composition,
giving another of 0.6~Gyr.  Oxygen abundance uncertainties of
$\sim$0.2~dex correspond to an age uncertainty of 0.6~Gyr.  These
systematics combine for a total uncertainty of 1.2~Gyr in absolute age.

\acknowledgements

Support for programs GO-13817, GO-14336, and GO-14773
was provided by NASA through a grant from
the Space Telescope Science Institute (STScI), which is operated by the
Association of Universities for Research in Astronomy, Inc., under
NASA contract NAS 5-26555.  
Based on observations obtained at the Southern Astrophysical Research
(SOAR) telescope, which is a joint project of the Minist\'{e}rio da
Ci\^{e}ncia, Tecnologia, Inova\c{c}\~{a}os e Comunica\c{c}\~{a}oes
(MCTIC) do Brasil, the U.S. National Optical Astronomy Observatory
(NOAO), the University of North Carolina at Chapel Hill (UNC), and
Michigan State University (MSU).
J.S. was supported by a Packard Fellowship.
We are grateful to B. Anthony-Twarog for
providing the full catalogs of her Str\"{o}mgren photometry.
The Digitized Sky Surveys were produced at STScI
under U.S. Government grant NAG W-2166. The images of these
surveys are based on photographic data obtained using the Oschin
Schmidt Telescope on Palomar Mountain and the UK Schmidt
Telescope. The plates were processed into the present compressed
digital form with the permission of these institutions.
The Second Epoch Survey of the southern sky was made by the
Anglo-Australian Observatory with the UK Schmidt Telescope.
Plates from this survey have been digitized and compressed by the STScI.
The digitized images are copyright 1993-2000 by the
Anglo-Australian Observatory Board, and are distributed herein by
agreement.

\end{document}